\input{aipcheck}

\documentclass[
    ,final            
  ]
  {aipproc}

\layoutstyle{6x9}

\def\Msolar{$\mathrm M_{\odot}$}

\begin{document}
\title{New population synthesis model \\  Preliminary results for close double white dwarf populations}


\author{Silvia Toonen}{
  address={Department of Astrophysics/IMAPP, Radboud University Nijmegen, the Netherlands} \thanks{e-mail: S. Toonen@astro.ru.nl}
}

\author{Gijs Nelemans}{
  address={Department of Astrophysics/IMAPP, Radboud University Nijmegen, the Netherlands} \thanks{e-mail: S. Toonen@astro.ru.nl}
}

\author{Simon F. Portegies Zwart}{
  address={Leiden Observatory, Leiden University, The Netherlands}
}

\begin{abstract}
An update is presented to the software package SeBa (\citet{Por96, Nel01}) for simulating single star and binary evolution in which new stellar evolution tracks (\citet{Hur00}) have been implemented. SeBa is applied to study the population of close double white dwarf and the delay time distribution of double white dwarf mergers that may lead to Supernovae Type Ia. 
\end{abstract}

\maketitle
\section{SeBa - a fast stellar \& binary evolution code}
\begin{figure}[!htb]
\begin{tabular}{c c}
\scalebox{0.325}{\includegraphics[trim=40mm 15mm 15mm 45mm, angle=270]{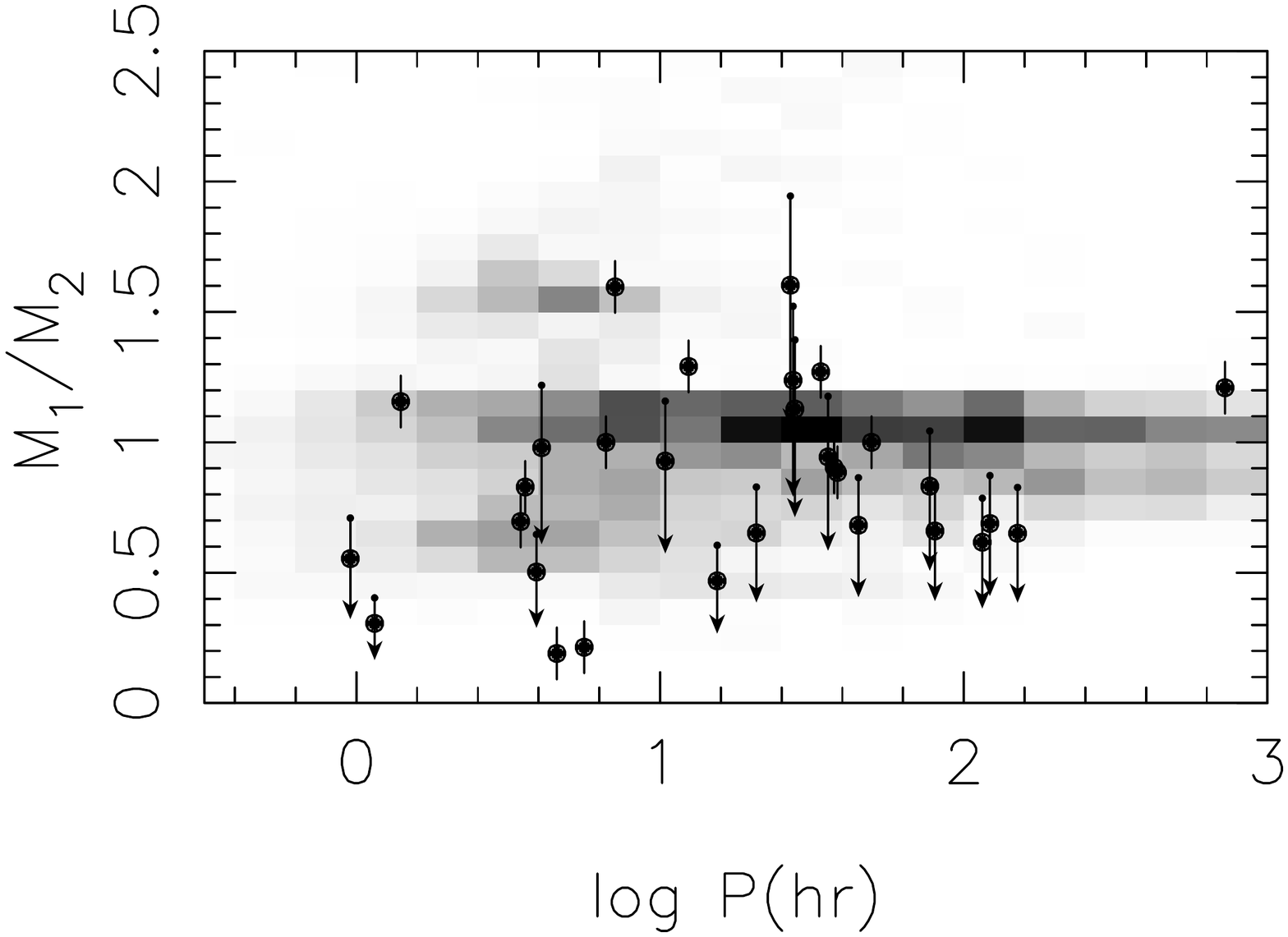}} &
\scalebox{0.325}{\includegraphics[trim=40mm 15mm 15mm 45mm, angle=270]{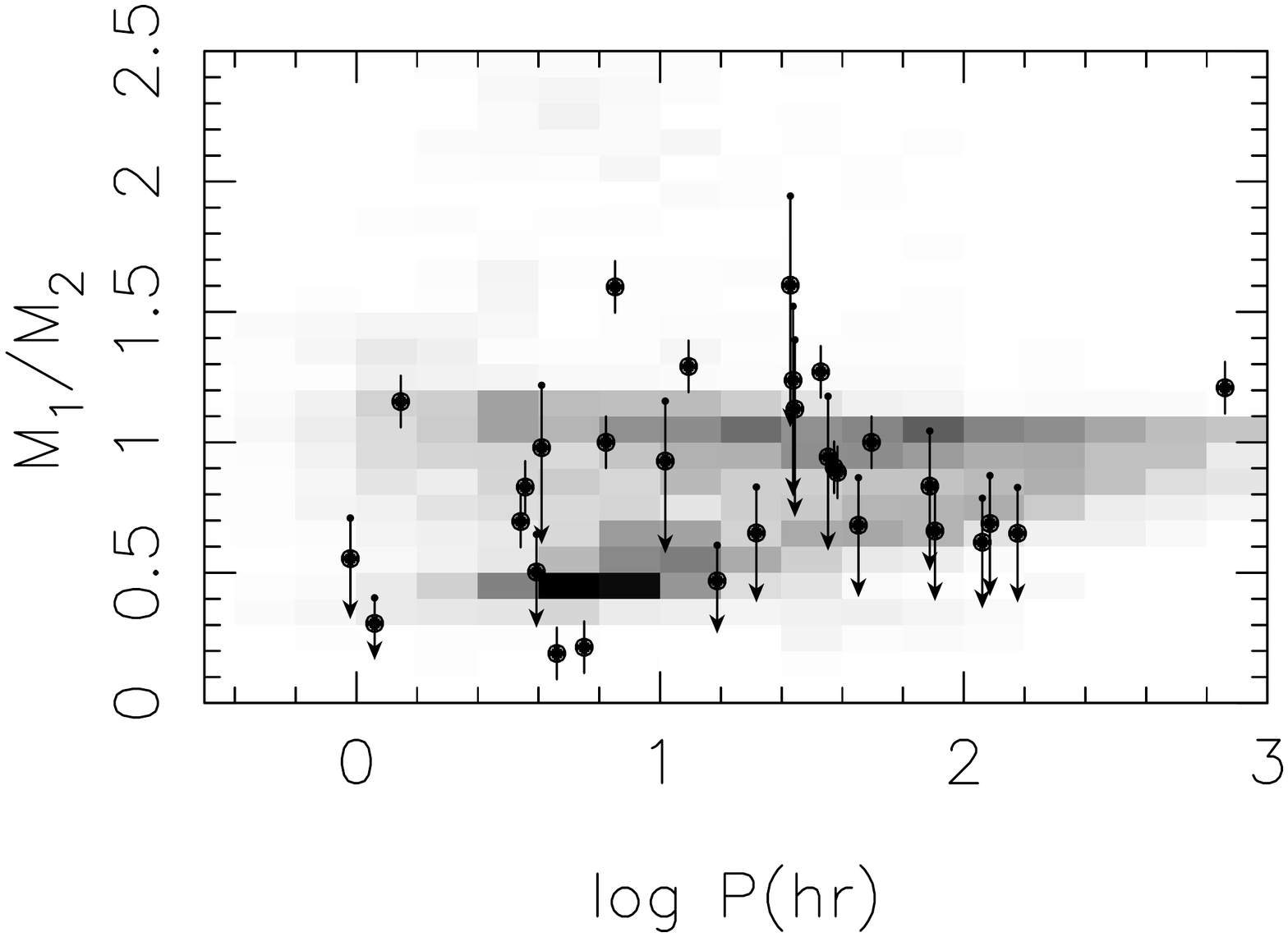}} \\
\small \emph{(a)} & \small \emph{(b)}\\
\end{tabular}
\caption{\emph{Simulated distribution of population of visible double white dwarfs as a function of orbital period and mass ratio, where mass ratio is defined as the mass of the brighter white dwarf devided by that of the dimmer white dwarf. Left the stellar evolution tracks according to EFT are used, right HPT. The intensity of the grey scale corresponds to the density of objects. Observed binary white dwarfs are overplotted with filled circles. }}
\label{fig:plaatjes1}
\end{figure}

We present an update to the software package SeBa (\citet{Por96, Nel01}) for simulating single star and binary evolution from the ZAMS until remnant formation including processes as mass transfer phases, common-envelope phases, magnetic braking and gravitational radiation. Previously stellar evolution has been based on evolutionary tracks described by analytic formulae given by \citet{Egg89} (hereafter EFT), but in the new version it is based on \citet{Hur00} (hereafter HPT). In this research SeBa is used to study close double white dwarfs. A comparison between simulations using the EFT and HPT stellar evolution tracks shows that the overall double white dwarf populations are similar, see Fig. \ref{fig:plaatjes1}. In the models presented here we assume solar metalicity and a 50\% binary fraction. Inital masses of single stars and binary primaries are distributed according to the Kroupa (\cite{Kro93}) IMF. Secondary masses are drawn from a flat mass ratio distribution. Initial orbital parameters are distributed according to a thermal eccentricity distribution and $\propto 1/a$ (\citet{Abt83}).  Differences between the populations are traced back to differences in the evolutionary tracks that affect the stability of mass transfer (at $M_1/M_2\approx1.5$ ) and the size of core masses of AGB stars (at $M_1/M_2<1$).   

\section{Delay times of double white dwarf mergers}
Close double white dwarfs tend to lose considerable amounts by the emission of gravitational radiation, which is not only interesting for future gravitational wave observatories as LISA (o.a. \citet{Eva87}), but also affects the binary system by decreasing the orbital period and possibly leading to a merger. The merger of two carbon-oxygen white dwarfs with a combined mass exceeding the Chandrasekhar mass limit is one of the possible scenarios for Type Ia Supernovae (SNe Ia). SNe Ia are very succesfully used as standard candles on cosmological distance scales, but so far the nature of the progenitor(s) is unclear.

The delay time distributions of these mergers for the EFT and the HPT stellar evolution tracks for solar metallicities are very similar. The delay time distribution for both types of tracks show that these mergers are expected to take place in young as well as old populations. 
The time-integrated number of SNe Ia per unit formed stellar mass is 4.3$\cdot$10$^{-4}$ \Msolar$^{-1}$ and 4.4$\cdot$10$^{-4}$ \Msolar$^{-1}$ when using the EFT resp. HPT tracks. From cluster SN Ia measurements, \citet{Mao10} infer a value of 5.9$\cdot$10$^{-3}$ \Msolar$^{-1}$ for their 'optimal' model. 
The current merger rate with a Galactic star formation rate of \citet{Boi99} is 1.4$\cdot$10$^{-3}$ yr$^{-1}$ for EFT as well as for HPT, where as the empirical Galactic SN Ia rate from \citet{Cap99} is (4$\pm$2)$\cdot$10$^{-3}$yr$^{-1}$.

\begin{figure}[!htb]
\begin{tabular}{c c}
\scalebox{0.36}{\includegraphics{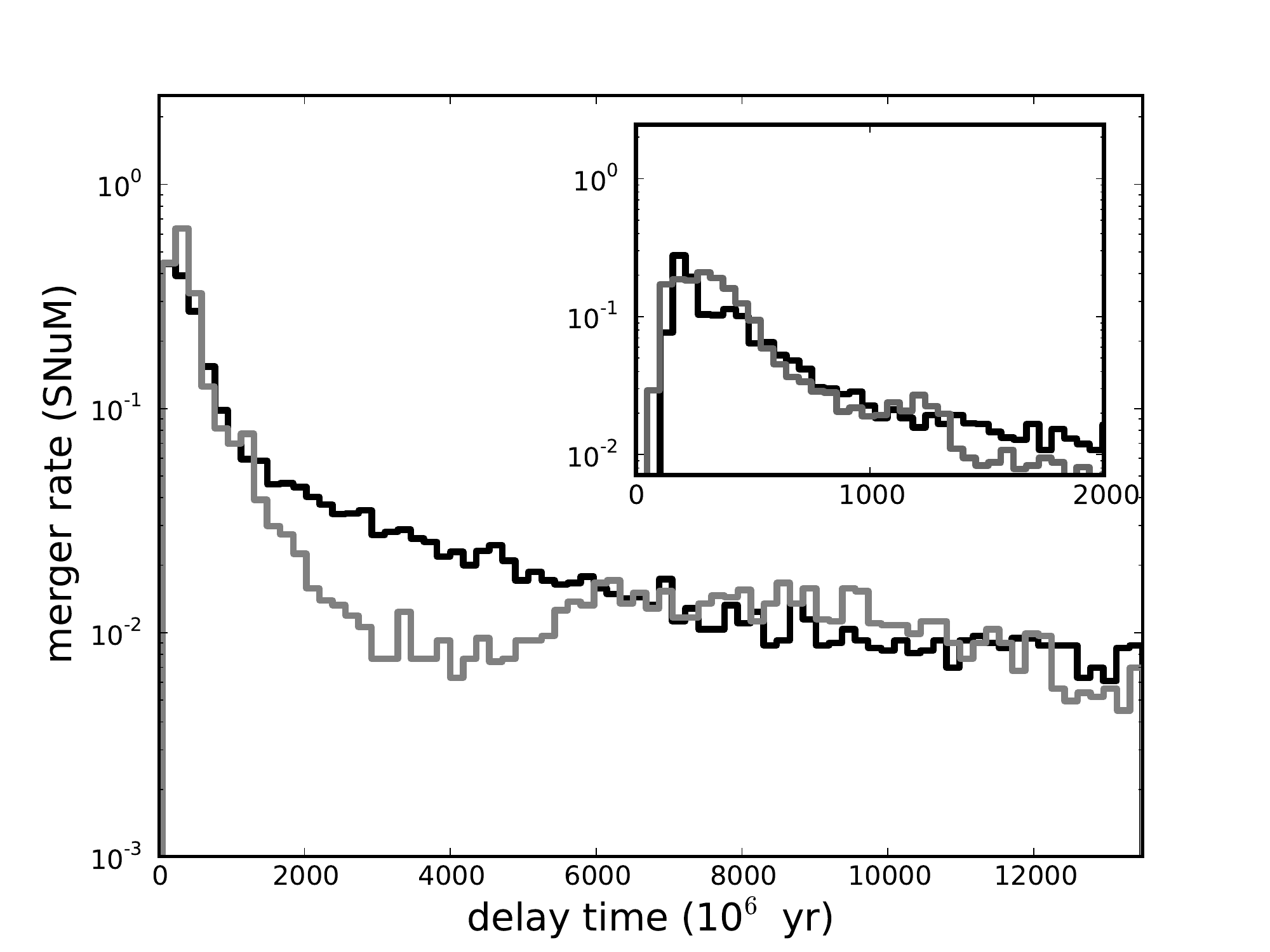}} &
\scalebox{0.36}{\includegraphics{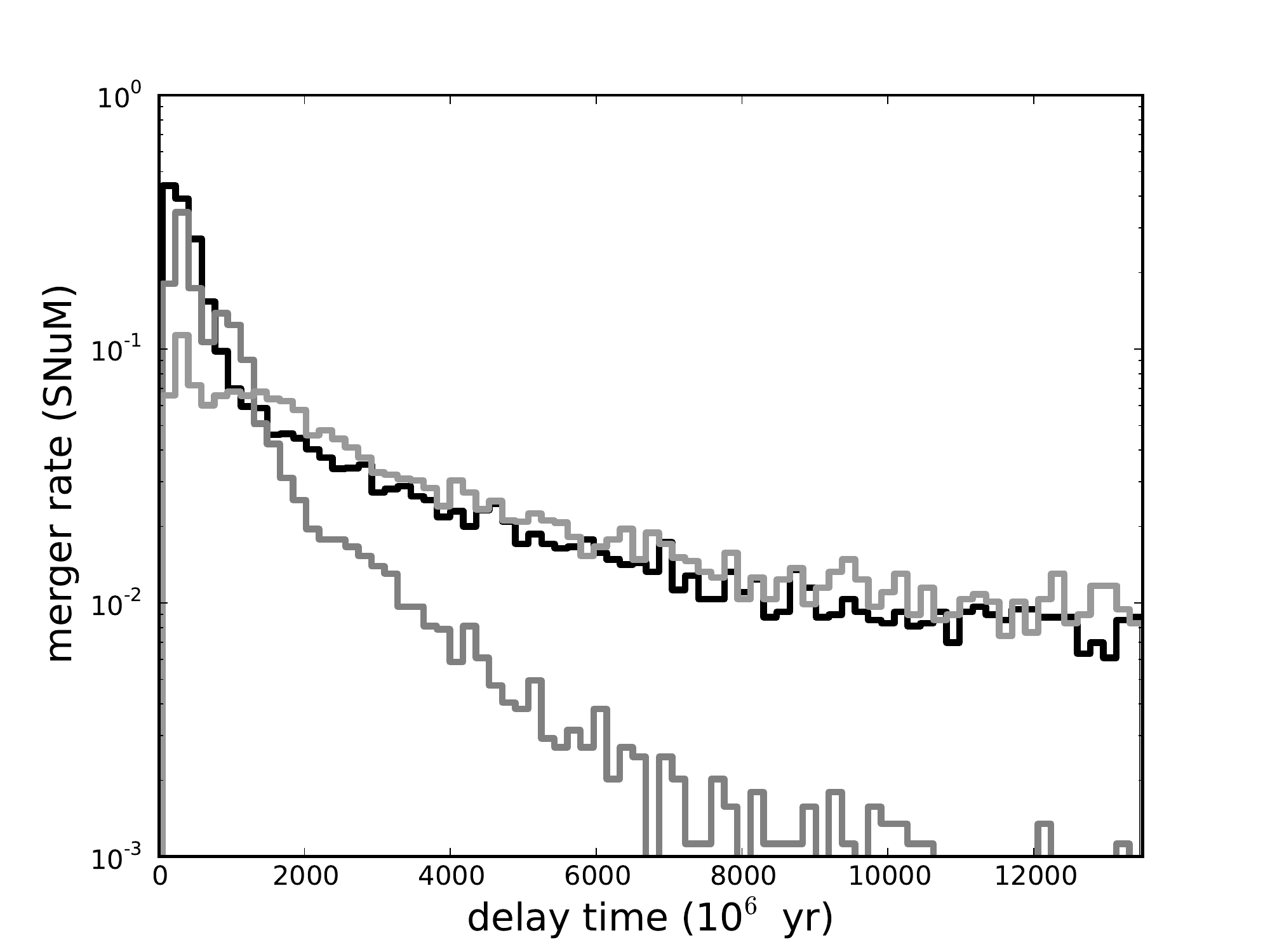}} \\
\small \emph{(a)} & \small \emph{(b)}\\
\end{tabular}
\caption{\emph{Merger rate of double carbon-oxygen white dwarfs with  total mass above the Chandrasekhar mass as a function of delay time. Rates are in (100 yr)$^{-1}$ per 10$^{10}$ \Msolar of the parent galaxy (SNuM). a) Black represents a simulation using the stellar evolution tracks according to HPT, grey EFT. b) Delay times using the HPT tracks are shown for three different prescriptions of the CE phase. In light-grey model $\alpha\alpha$, in grey model $\gamma\gamma$ and in black model $\gamma\alpha$. }}
\label{fig:plaatjes2}
\end{figure}

\newpage
\section{Common envelope evolution}
Close double white dwarfs are believed to encounter at least two phases of common envelope (CE) evolution. In spite of the importance of the CE phase, it remains poorly understood. Several prescriptions for CE evolution have been proposed. The $\alpha$ formalism (\citet{Web84}) is based on the conservation of orbital energy The $\alpha$-parameter describes the efficiency of which orbital energy is consumed to unbind the common envelope according to:
\begin{equation}
E_{gr} = \alpha (E_{orb,init}-E_{orb,final})
\end{equation}
where $E_{orb}$ the orbital energy and $E_{gr}$ is the binding energy  between the envelope mass $M_{1,env}$ and the mass of the primary $M_1$. $E_{gr}$ is often approximated by:
\begin{equation}
E_{gr} = \frac{GM_1 M_{1,env}}{\lambda R}
\end{equation} 
where $R$ is the radius of the primary and $\lambda$ depends on the structure of the primary star. We assume $\alpha \lambda = 2$. 

In order to explain to the observed distribution of double white dwarfs \citet{Nel00} proposed an alternative formalism. According to this $\gamma$ formalism, mass transfer is unstable and non-conservative. The mass loss reduces the angular momentum of the system in a linear way according to:
\begin{equation}
\frac{J_{init}-J_{final}}{J_{init}} = \gamma \frac{\Delta M}{M_1+ M_2}
\end{equation} 
where $J_{init}$ resp. $J_{final}$ is the angular momentum of the pre- respectively post-mass transfer binary and $M_2$ is the mass of the secondary. We assume a value of 1.75 for $\gamma$.

In Fig. \ref{fig:plaatjes2} we compare the delay time distribution for three different models of CE evolution. In model $\alpha\alpha$ the $\alpha$ formalism is used to determine the outcome of every CE, in model $\gamma\gamma$ the $\gamma$ formalism is used every time. The preferred model $\gamma\alpha$ is a  combination of the $\alpha$ and $\gamma$ formalisms as in \citet{Nel00}. In systems with a giant primary and  remnant or giant companion the CE phase is according to $\alpha$, for a giant primary and a non-remnant or non-giant  $\gamma$ is used. The $\gamma\alpha$ model shows high rates at all delay times, where as the rate for model $\alpha\alpha$ decreases significantly at long delay times and for $\gamma\gamma$ saturates at a level below 0.5 (100 yr)$^{-1}$ per 10$^{10}$ \Msolar at short delay times. As a result the current Galactic merger rates of model $\alpha\alpha$ (5.9$\cdot$10$^{-4}$ yr$^{-1}$) and $\gamma\gamma$ (1.2$\cdot$10$^{-3}$ yr$^{-1}$) are $\sim$ 40\% and $\sim$90\% respectively of that of model $\gamma\alpha$ (1.4$\cdot$10$^{-3}$ yr$^{-1}$).  The time-integrated number of SNe Ia per unit formed stellar mass is 4.4$\cdot$10$^{-4}$ \Msolar$^{-1}$, 2.8$\cdot$10$^{-4}$ \Msolar$^{-1}$ and 3.3$\cdot$10$^{-4}$ \Msolar$^{-1}$ for model $\gamma\alpha$, $\alpha\alpha$ and $\gamma\gamma$ respectively.

\begin{theacknowledgments}
We wish to thank the organizers for a stimulating and enjoyable conference and congratulate Ron Webbink with his 65$^{\mathrm{th}}$ birthday as well as Peter Eggleton and Ed van de Heuvel with their birthdays.  
\end{theacknowledgments}

\bibliographystyle{aipproc}
\bibliography{bibtex_silvia_toonen}

\end{document}